\documentclass[twocolumn,showpacs,preprintnumbers,amsmath,amssymb]{revtex4-1}
\usepackage{graphicx}

\begin{document}

\title{Improving the quality of noisy spatial quantum channels}
\author{Ning Tang, Zi-Long Fan and Hao-Sheng Zeng}
\email{Corresponding author: hszeng@hunnu.edu.cn}
\affiliation{Key Laboratory of Low-Dimensional Quantum
Structures and Quantum Control of Ministry of Education, and
Department of Physics, Hunan Normal University, Changsha 410081,
China}

\begin{abstract}
We show, for the non-Markovian or time-dependent Markovian model of noise, by breaking the noisy spatial quantum channel (SQC) into a series of periodically arranged sub-components,
that the quality of information transmission described by the purity, fidelity and concurrence of the output states can be improved. The relation to the quantum Zeno effect and possible applications of the idea have been discussed.

\end{abstract}

\maketitle

\section{Introduction}
Storage and transmission of quantum information are the foundation for quantum information processing. The former processes correspond to time quantum channels (TQCs), while the later one are fulfilled through SQCs. Due to the inevitable interaction of quantum systems with their environments, both TQCs and SQCs are noisy. For noisy TQCs, the stored quantum information (or quantum memory) touches or interacts with the same environment, which thus can be described by means of the normal open-quantum-system dynamics \cite{Breuer}.

SQCs are the pipelines for quantum information transmission which are useful in many tasks of quantum communication, such as quantum teleportation \cite{Bennett}, quantum cryptography \cite{Ekert} and distant entanglement distribution \cite{Cirac,Holloway}. During the transmission, the quantum entity that carries quantum information at different times locates different positions of the pipeline and thus possibly interacts with different environments. These environments could be independent or correlated each other, depending on the property of the pipeline. Thus the transmission of quantum information through a noisy SQC should be described by a series of successively-proceeding open quantum dynamics, each with correlated or independent environments. This makes the treatment of noisy SQCs much more sophisticated than that of noisy TQCs. In a special situation where the pipeline is made up of a series of independent segments, each regarded as an independent local environment, then the process of information transmission over the whole SQC can be modeled by a series of successive but independent open dynamics. This special kind of noisy SQCs Make up actually the foundation of our study.

The evolution of open quantum systems may be distinguished into two basic types: Markovian and non-Markovian processes.
For the Markovian process, the correlation time between the system and
environment is considered to be infinitesimally small so that the
dynamical map does not carry any memory effects, leading to
the monotonic flow of information \cite{Breuer1,Laine} or the divisible dynamics \cite{Rivas} of the open system. According to the recent investigation for open quantum systems, Markovian processes also include two forms \cite{Alicki}: time-independent and time-dependent. If a quantum process is governed by the standard Lindblad master equation \cite{Gorini,Lindblad}, the corresponding dynamics is of course Markovian. Due to the time independence of the generator, this dynamical process is often called time-independent Markovian process. By contrast, if a dynamics is governed by the Lindblad-like master equation where the generator is time-dependent and the decay rates are positive all the time, the corresponding dynamics is also Markovian and is known as time-dependent Markovian process. In this paper, we call the noises that correspond to the two types of Markovian processes respectively the time-independent and time-dependent Markovian model of noises. As we will see in the text, these two types of Markovian model of noises can make the SQCs to have distinct features (The time-dependent Markovian model of noises behave like non-Markovian noises in the aspect of impacting the quality of quantum information transmission), though they all lead to Markovian dynamics.

For the non-Markovian process, the memory effect will give rise to, besides the backflow of information \cite{Breuer1,Laine}, many new dynamical traits such as the non-monotonic reduction of entanglement \cite{Rivas} or correlation \cite{Luo}, the correlation quantum beat \cite{Zeng}, the entanglement trapping \cite{Bellomo}, and the phenomenon of impossible thermalization \cite{Wang}, etc. Furthermore, the non-Markovianity may be used as useful resources in the context of quantum metrology \cite{Chin1}, quantum key distribution \cite{Vasile}, quantum teleportation \cite{Laine2}, and improving the quantum capacity \cite{Bylicka}. In this paper, we show that the non-Markovian effect can also be used to improve the quality of quantum information transmission. In the demonstration, we only involve that kind of non-Markovian noises where the master equation has the time-local Lindblad-like form with temporarily negative decay rates. Recent research \cite{Maldonado,Laine1} suggests that this kind of master equation can describe quite a number of realistic non-Markovian processes. 

The paper is organized as follows. In Sec. II, we present the model of the toy noisy SQC and the physical quantities for describing the quality of the noisy SQC. In Sec. III and Sec. IV, we study respectively the properties of phase-damping and amplitude-damping SQCs. And Sec. V is designed for the illustration of the Markovian limit and quantum Zeno effect involved in our work. The conclusion is arranged in Sec. VI.

\section{Model for a noisy SQC}
We begin with our investigation by engineering a toy noisy SQC which is formed by a series of periodically arranged cavities (Fig.1). A two-level atom (qubit) passes from the left through all the cavities. Assuming in each cavity the atom stays a time $\Delta t$ and the time between cavities can be ignored, thus the total time the atom spends over the SQC is $T=n\Delta t$. Assume that the properties of each cavity are completely same, so that in each cavity the atom evolves via an identical Hamiltonian $H$. According to the operator-sum representation of quantum operations \cite{Nielsen}, the evolution for the atom in each cavity can be described by the map
\begin{equation}
    \varepsilon(\rho)=\sum_{j}K_{j}(t)\rho K_{j}^{+}(t),
\end{equation}
where the Kraus operators $K_{j}(t)$ fulfill the trace-preserving condition $\sum_{j}K_{j}^{+}(t)K_{j}(t)=I$. After over the whole SQC, the atom state is thus given by
\begin{equation}
    \rho(T)=\varepsilon_{n}\bigg(\varepsilon_{n-1}\Big(\cdots\varepsilon_{1}\big(\rho(0)\big)\cdots\Big)\bigg),
\end{equation}
which means that the atom evolves from the initial state $\rho(0)$, with the output state of the former map as the input of the later map. The action time of each map is $\Delta t$.

In this paper, we will investigate the properties of the noisy SQC in terms of the following quantities. As everyone knows, for a two-level system, the non-diagonal element $\rho_{01}$ of the density matrix denotes the quantum coherence of the given quantum state. The non-diagonal element reaches its maximum for the pure quantum superposition state, and reduces after decoherence. Thus we define by the value of the non-diagonal element of the density matrix of the atomic output state,
\begin{equation}
    P=|\rho_{01}(T)|,
\end{equation}
the purity of the atomic output state, which describes the coherence-preserving ability of the atomic states after over the noisy SQC. The second quantity is the fidelity defined by the overlap of the output state $\rho(T)$ with the input state $\rho(0)$
\begin{equation}
    F=tr\sqrt{\sqrt{\rho(0)}\rho(T)\sqrt{\rho(0)}},
\end{equation}
which describes the state-preserving ability for transmitting quantum states through the noisy SQC. The last quantity is used to describe the entanglement-preserving ability for entanglement distribution through the noisy SQC. Suppose we have a two-qubit locally entangled state $\rho_{12}(0)$ and wish to establish remote entanglement from it, we thus let one of the qubit transmit to the remote receiver via the SQC. After this, a nonlocal entanglement state $\rho_{12}(T)$ between the two qubits is formed, whose entanglement may be described by the well-known concurrence \cite{Wootters1998}
\begin{equation}
C=\max\{0,\lambda_{1}-\lambda_{2}-\lambda_{3}-\lambda_{4}\}.
\end{equation}
Here $\lambda_{1}\geq\lambda_{2}\geq\lambda_{3}\geq\lambda_{4}$ are
the square roots of the eigenvalues of the matrix $R=\rho_{12}(T)
\widetilde{\rho}_{12}(T)$, with
$\widetilde{\rho}_{12}(T)=\sigma_{y}^{(1)}\otimes\sigma_{y}^{(2)}\rho^{*}_{12}(T)\sigma_{y}^{(1)}\otimes\sigma_{y}^{(2)}$
and the sign ``$\ast$" standing for complex conjugate. The larger the concurrence is, the better the ability for the preservation of entanglement will be.

At the end of this section, we would like to make mention of the rationality of the toy SQC. The elements in the model may be replaced in practice. The two-level atom is just
an indicator of a qubit which may actually be any a two-state system, such as a photon with two independent polarizations. Correspondingly,
the SQC may be constituted by a series of optical fiber segments instead of the cavities. If the gap between neighboring fiber segments is much less than the length of each fiber segment, then the time for the photon between gaps can be neglected. The toy SQC thus becomes realistic in practice.

\begin{figure}[htp] \center
\includegraphics[scale=0.6]{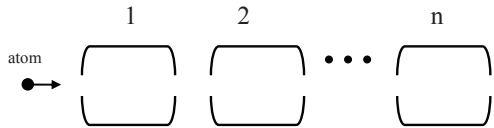}
\caption{A toy noisy SQC formed by $n$ cavities is passed by a two-level atom. Denoted by $\Delta t$ the time the atom stays in each cavity, the total time over the whole SQC is thus $T=n\Delta t$.}
\end{figure}

\section{Dephasing SQC}
Now let us proceed in detail with the case of dephasing SQC. The Hamiltonian for the interaction of the two-level atom with each cavity is given by
\begin{equation}
    H=\frac{\omega_{0}}{2}\sigma_{z}+\sum_{k}\omega_{k}b^{+}_{k}b_{k}+\sum_{k}\sigma_{z}(\lambda_{k}b^{+}_{k}+\lambda^{*}_{k}b_{k}).
\end{equation}
Where $\omega_{0}$ and
$\sigma_{z}$ are respectively the transition frequency and Pauli
operator of the atom, and $\omega_{k}$, $b_{k}$ are respectively the
frequency and annihilation operator for the $k$-th harmonic
oscillator of the reservoir. The coupling strength $\lambda_{k}$ is
assumed to be complex in general. This dephasing model, which is
extensively used to simulate the decoherence of a qubit coupled to
its environment in quantum information science, can be solved
exactly. For the initial product states of the qubit plus its environment, the evolution
of the reduced density matrix of the qubit, in the basis $\{|0\rangle, |1\rangle\}$, may be
written as \cite{Breuer}
\begin{equation}
    \rho(t)=\left(
              \begin{array}{cc}
                \rho_{00} & \rho_{01}e^{-\Gamma(t)} \\
                \rho_{10}e^{-\Gamma(t)} & \rho_{11} \\
              \end{array}
            \right).
\end{equation}
Where the decoherence function is defined by
\begin{equation}
    \Gamma(t)=-\sum_{k}\ln\langle\exp[\alpha_{k}b^{+}_{k}-\alpha^{*}_{k}b_{k}]\rangle,
\end{equation}
with $\alpha_{k}=2\lambda_{k}(1-e^{-i\omega_{k}t})/\omega_{k}$, and the angular brackets denoting the expectation value with respect to the bath state. The evolution of Eq.(7) satisfies master equation
\begin{equation}
    \frac{d\rho}{dt}=\frac{1}{2}\gamma_{p}(t)[\sigma_{z}\rho\sigma_{z}-\rho],
\end{equation}
with dephasing rate $\gamma_{p}(t)=\dot{\Gamma}(t)$.
In the operator-sum representation, one gets two Kraus operators
\begin{equation}
    K_{1}(t)=\sqrt{\frac{1+e^{-\Gamma(t)}}{2}}I, K_{2}(t)=\sqrt{\frac{1-e^{-\Gamma(t)}}{2}}\sigma_{z}.
\end{equation}

Now for the case under consideration, the atom passes over the whole SQC which includes $n$ completely identical cavities,  thus the output state of the atom
can be evaluated from Eq.(2) as
\begin{equation}
\rho(T)=\left(
              \begin{array}{cc}
                \rho_{00} & \rho_{01}e^{-T\frac{\Gamma(\Delta t)}{\Delta t}} \\
                \rho_{10}e^{-T\frac{\Gamma(\Delta t)}{\Delta t}} & \rho_{11} \\
              \end{array}
            \right).
\end{equation}
It shows that the coherence decays exponentially with the total time $T$, with the decay rate given by $R(\Delta t)\equiv\Gamma(\Delta t)/\Delta t$. In the following, we will show through exemplary examples that $R(\Delta t)$ may be reduced by decreasing $\Delta t$.
For this dephasing SQC, the purity and the output-to-input fidelity are
\begin{equation}
    P=|\rho_{01}|e^{-T\frac{\Gamma(\Delta t)}{\Delta t}},
\end{equation}
and
\begin{eqnarray}
    F&=&\rho^{2}_{00}+\rho^{2}_{11}+2\rho_{01}\rho_{10}e^{-T\frac{\Gamma(\Delta t)}{\Delta t}}\\ \nonumber &+&2\sqrt{(\det\rho)[\rho_{00}\rho_{11}-\rho_{01}\rho_{10}e^{-T\frac{\Gamma(\Delta t)}{\Delta t}}]},
\end{eqnarray}
respectively, where $\det\rho=\rho_{00}\rho_{11}-\rho_{01}\rho_{10}$ is the determinant of $\rho$. In order to discuss the entanglement-preserving ability of the SQC, we assume the initial local entanglement state of the two qubits is an arbitrary pure state
\begin{equation}
    |\Psi(0)\rangle=a|00\rangle+b|01\rangle+c|10\rangle+d|11\rangle,
\end{equation}
with the complex superposition coefficients satisfying normalization $|a|^{2}+|b|^{2}+|c|^{2}+|d|^{2}=1.$ Note that this pure state has concurrence $C(0)=2|ad-bc|$. After the first qubit transmits through the SQC to another receiver, the nonlocal entanglement state becomes
\begin{equation}
    \rho_{\Psi}(T)=\left(
              \begin{array}{cccc}
                |a|^{2} & ab^{*} & ac^{*}E(T) & ad^{*}E(T) \\
                a^{*}b & |b|^{2} & bc^{*}E(T) & bd^{*}E(T) \\
                a^{*}cE(T) & b^{*}cE(T) & |c|^{2} & cd^{*} \\
                a^{*}dE(T) & b^{*}dE(T) & c^{*}d & |d|^{2} \\
              \end{array}
            \right)
\end{equation}
with $E(T)=e^{-T\frac{\Gamma(\Delta t)}{\Delta t}}$, and the corresponding concurrence is
\begin{equation}
    C=2|ad-bc|e^{-T\frac{\Gamma(\Delta t)}{\Delta t}}.
\end{equation}
Eqs.(12) and (16) show that both the purity and the concurrence take on simple role of exponentially decay. The normalized quantities, $P(T)/P(0)$ and $C(T)/C(0)$, are same and independent of the initial states. While the fidelity of Eq.(13) has a complicated rule of decay which depends on the initial states of the atom.

The decoherence function $\Gamma(t)$ of Eq. (8) depends on the properties of environments. In the following, we will consider two kinds of commonly encountered baths:
thermal bath and squeezed vacuum bath. For the thermal bath, the decoherence function becomes \cite{Breuer}
\begin{equation}
    \Gamma(t)=\int_{0}^{\infty}d\omega J(\omega)\coth(\frac{\omega}{2k_{B}T_{B}})\frac{1-\cos(\omega t)}{\omega^{2}},
\end{equation}
with $k_{B}$ the
Boltzmann constant and $T_{B}$ the bath temperature. The spectral density is defined as $J(\omega)=\sum_{k}4|\lambda_{k}|^{2}\delta(\omega-\omega_{k})$. For the squeezed vacuum bath $\rho_{B}=|\phi\rangle\langle\phi|$ with $|\phi\rangle=\prod_{k}S_{k}(\xi_{k})|0\rangle$ and unitary squeeze operator $S_{k}(\xi_{k})=\exp[\frac{1}{2}\xi^{*}_{k}b^{2}_{k}-\frac{1}{2}\xi_{k}b^{+2}_{k}]$, the decoherence function becomes
\begin{eqnarray}
    \Gamma(t)&=&-\int_{0}^{\infty}d\omega J(\omega)\frac{1-\cos(\omega t)}{\omega^{2}}
    \\
    \nonumber &\times&\left\{\cosh[2r(\omega)]-\sinh[2r(\omega)]\cos[\omega t-\theta(\omega)]\right\},
\end{eqnarray}
where $r(\omega)$ and $\theta(\omega)$ are respectively the amplitude and argument of the squeeze parameter $\xi_{k}$.
\begin{figure}[htp] \center
\includegraphics[scale=0.6]{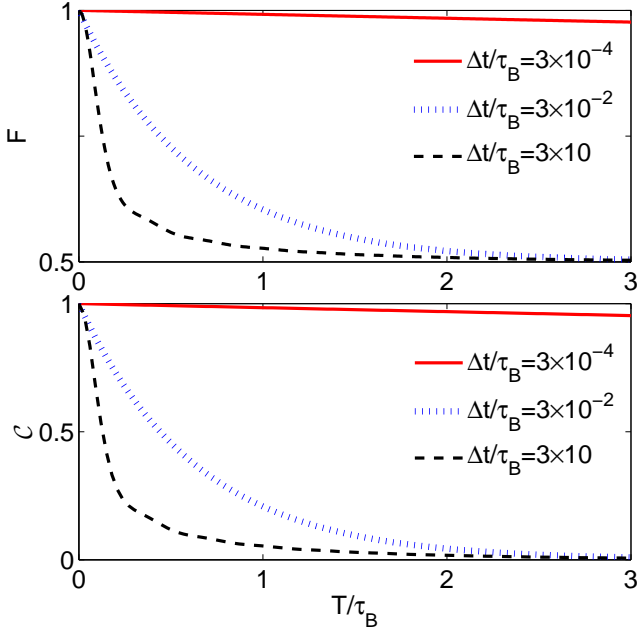}
\caption{(Color online) Evolutions of the fidelity and the normalized concurrence versus the dimensionless time $T/\tau_{B}$ for thermal bath and Ohmic spectrum $J(\omega)=\omega e^{-\omega/\Omega}$ with $\Omega\tau_{B}=20$. The fidelity is calculated under the initial state $|\varphi(0)\rangle=(|0\rangle+|1\rangle)/\sqrt{2}$ of the atom.}
\end{figure}

In Fig.2, we show, for the thermal bath and Ohmic spectrum $J(\omega)=\omega e^{-\omega/\Omega}$, the evolutions of the fidelity and the normalized concurrence versus total time $T$ for three kinds of $\Delta t$. Where the dimensionless time is realized through the thermal correlation time $\tau_{B}=1/k_{B}T_{B}$ and we take $\Omega\tau_{B}=20$. The evolution of normalized purity is the same as that of the normalized concurrence and thus not shown. In the calculation of fidelity, we take the initial state of the atom to be $|\varphi(0)\rangle=(|0\rangle+|1\rangle)/\sqrt{2}$. We see, for given $\Delta t$, that both the fidelity and concurrence decrease with the total time $T$, which agree with our intuitive senses. However, when the total time $T$ is fixed, the fidelity and concurrence increase with the decreasing of $\Delta t$. In the limit of $\Delta t\rightarrow 0$, we have from Eq.(17) that the decay rate $\Gamma(\Delta t)/\Delta t\rightarrow 0$, so that the fidelity and the normalized concurrence approach to unity. The output states of Eqs.(11) and (15) also return respectively to their initial states.

\begin{figure}[htp] \center
\includegraphics[scale=0.6]{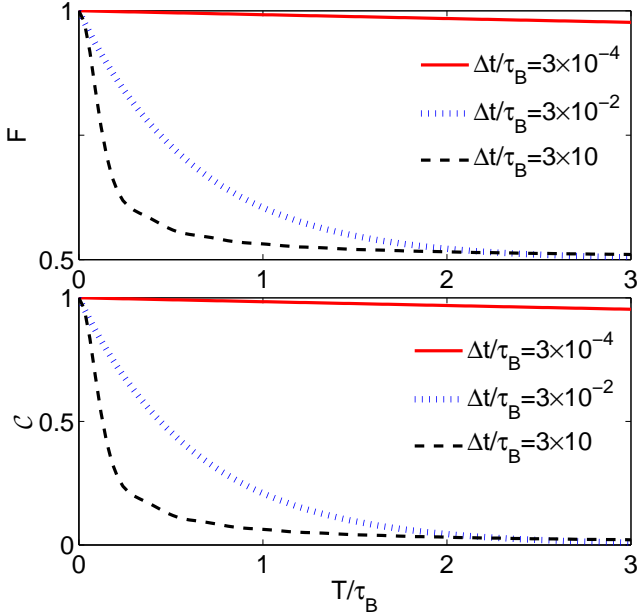}
\caption{(Color online) Evolutions of the fidelity and the normalized concurrence versus the dimensionless time $T/\tau_{B}$ for squeezed vacuum bath and Ohmic spectrum $J(\omega)=\omega e^{-\omega/\Omega}$ with $\Omega\tau_{B}=20$. The fidelity is calculated under the initial state $|\varphi(0)\rangle=(|0\rangle+|1\rangle)/\sqrt{2}$ of the atom.}
\end{figure}

Fig.3 shows, for the squeezed vacuum bath and Ohmic spectrum $J(\omega)=\omega e^{-\omega/\Omega}$, the evolutions of the fidelity and the normalized concurrence versus total time $T$ for three kinds of $\Delta t$. Where we also take $\Omega\tau_{B}=20$ and the initial state of the atom to be $|\varphi(0)\rangle=(|0\rangle+|1\rangle)/\sqrt{2}$ for the calculation of fidelity. In addition, we take the mode-dependent squeezed amplitude to be a Gaussian distribution $$r(\omega)=\frac{r_{0}}{\sqrt{2\pi}\sigma}e^{-\frac{(\omega-\omega_{0})^{2}}{2\sigma^{2}}},$$ with $r_{0}=3$, $\omega_{0}=10^{9}\texttt{Hz}$ and width $\sigma=100\texttt{Hz}$. We assume $\theta(\omega)\equiv\pi/4$ which is independent of modes. The figure also shows the similar properties as that of Fig.2. In the limit of $\Delta t\rightarrow 0$, we also have from Eq.(18) that the decay rate $\Gamma(\Delta t)/\Delta t\rightarrow 0$, so that the fidelity and the normalized concurrence approach to unity.

\section{Amplitude-damping SQC}
In the previous section, we discussed the situation where the noise of the SQC is the purely dephasing type. Another typical noise is the so-called dissipation,
\begin{equation}
    H=\omega_{0}\sigma_{z}+\sum_{k}\omega_{k}b^{+}_{k}b_{k}+\sum_{k}(g_{k}b_{k}\sigma_{+}+g^{*}_{k}b^{+}_{k}\sigma_{-}),
\end{equation}
with $\sigma_{+}$, $\sigma_{-}$ the ladder operators of the atom, and $g_{k}$ the coupling constant of the atom with the bath mode $k$. All other parameters are the same as before. For the initial vacuum state of environment, this model can be solved exactly which gives the evolution of the reduced density of the atom \cite{Laine},
\begin{equation}
    \rho(t)=\left(
              \begin{array}{cc}
                1-|G(t)|^{2}\rho_{11} & G(t)\rho_{01} \\
                G^{*}(t)\rho_{10} & |G(t)|^{2}\rho_{11} \\
              \end{array}
            \right),
\end{equation}
where the function $G(t)$ fulfils the integro-differential equation
\begin{equation}
    \dot{G}(t)=-\int_{0}^{t}dt_{1}f(t-t_{1})G(t_{1}),
\end{equation}
with initial condition $G(0)=1$. The two-point reservoir correlation function $f(t-t_{1})$
relates to the spectral density $J(\omega)$ via the Fourier transformation $f(t-t_{1})=\int d\omega J(\omega)\exp[i(\omega_{0}-\omega)(t-t_{1})]$ with $\omega_{0}$ the transition frequency of the atom. The dynamical state of Eq.(20) satisfies the time-local master equation
\begin{equation}
    \frac{d\rho}{dt}=-\frac{i}{2}S(t)[\sigma_{+}\sigma_{-},\rho]+\gamma_{a}(t)\left[\sigma_{-}\rho\sigma_{+}-\frac{1}{2}\{\sigma_{+}\sigma_{-},\rho\}\right],
\end{equation}
with
\begin{equation}
    S(t)=-2\texttt{Im}\left[\frac{\dot{G}(t)}{G(t)}\right], \gamma_{a}(t)=-2\texttt{Re}\left[\frac{\dot{G}(t)}{G(t)}\right].
\end{equation}
In operator-sum representation, there are accordingly two Kraus operators,
\begin{equation}
    K_{1}(t)=\left(
               \begin{array}{cc}
                 1 & 0 \\
                 0 & G^{*}(t) \\
               \end{array}
             \right)
    , K_{2}(t)=\left(
                 \begin{array}{cc}
                   0 & \sqrt{1-|G(t)|^{2}} \\
                   0 & 0 \\
                 \end{array}
               \right).
\end{equation}
After the atom passes over the whole SQC, the output state becomes
 \begin{equation}
    \rho(T)=\left(
              \begin{array}{cc}
                1-\rho_{11}|G(\Delta t)|^{\frac{2T}{\Delta t}} & \rho_{01}[G(\Delta t)]^{\frac{T}{\Delta t}} \\
                \rho_{10}[G^{*}(\Delta t)]^{\frac{T}{\Delta t}} & \rho_{11}|G(\Delta t)|^{\frac{2T}{\Delta t}} \\
              \end{array}
            \right).
\end{equation}
The purity and the output-to-input fidelity are thus
\begin{equation}
    P=|\rho_{01}|\cdot |G(\Delta t)|^{\frac{T}{\Delta t}},
\end{equation}
and
\begin{eqnarray}
    F&=&\rho_{00}+\rho_{11}(\rho_{11}-\rho_{00})|\mathbb{G}|^{2}+2\mathrm{Re}[\rho_{01}\rho_{10}\mathbb{G}]
    \\ \nonumber
    &+&2\sqrt{(\det\rho)[\rho_{11}|\mathbb{G}|^{2}(1-\rho_{11}|\mathbb{G}|^{2})-\rho_{01}\rho_{10}|\mathbb{G}|^{2}]},
\end{eqnarray}
respectively, where $\mathbb{G}=[G(\Delta t)]^{T/\Delta t}$. For the entanglement distribution, we still assume the local initial state of Eq.(14) and let the first atom over the SQC, then the resultant nonlocal two-atom state becomes
\begin{widetext}
\begin{equation}
    \rho_{\Psi}(T)=\left(
              \begin{array}{cccc}
                |a|^{2}+|c|^{2}(1-|\mathbb{G}|^{2}) & ab^{*}+cd^{*}(1-|\mathbb{G}|^{2}) & ac^{*}\mathbb{G} & ad^{*}\mathbb{G} \\
                a^{*}b+c^{*}d(1-|\mathbb{G}|^{2}) & |b|^{2}+|d|^{2}(1-|\mathbb{G}|^{2}) & bc^{*}\mathbb{G} & bd^{*}\mathbb{G} \\
                a^{*}c\mathbb{G}^{*} & b^{*}c\mathbb{G}^{*} & |c|^{2}|\mathbb{G}|^{2} & cd^{*}|\mathbb{G}|^{2} \\
                a^{*}d\mathbb{G}^{*} & b^{*}d\mathbb{G}^{*} & c^{*}d|\mathbb{G}|^{2} & |d|^{2}|\mathbb{G}|^{2} \\
              \end{array}
            \right),
\end{equation}
\end{widetext}
which has the concurrence,
\begin{equation}
    C=2|ad-bc|\cdot|G(\Delta t)|^{\frac{T}{\Delta t}}.
\end{equation}

For the amplitude-damping SQC, we only discuss the case where the spectral density of each cavity is Lorentzian
\begin{equation}
    J(\omega)=\frac{\gamma_{0}\lambda^{2}}{2\pi[(\omega_{0}-\omega-\Delta)^{2}+\lambda^{2}]},
\end{equation}
where $\Delta=\omega_{0}-\omega_{c}$ is the detuning between the atomic frequency $\omega_{0}$ and cavity-mode $\omega_{c}$, $\gamma_{0}$ and
$\lambda$ are respectively the atomic free decay rate and the photon-leakage rate of the cavity. For this Lorentzian structured environment, the function $G(t)$ in Eq. (21) may be evaluated as
\begin{equation}
    G(t)=e^{-(\lambda-i\Delta)t/2}\left[\cosh(\frac{\delta t}{2})+\frac{\lambda-i\Delta}{\delta}\sinh(\frac{\delta t}{2})\right],
\end{equation}
with $\delta=\sqrt{(\lambda-i\Delta)^{2}-2\gamma_{0}\lambda}$.

\begin{figure}[htp] \center
\includegraphics[scale=0.6]{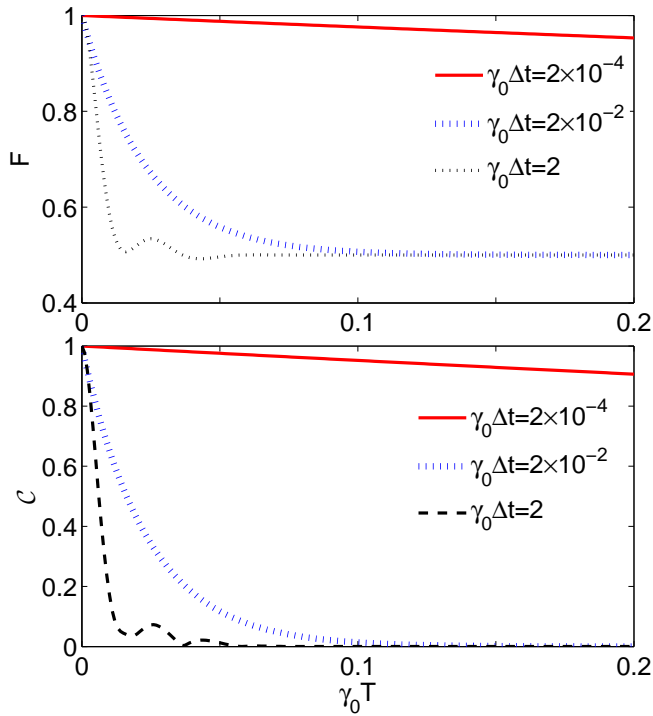}
\caption{(Color online) Evolutions of the fidelity and the normalized concurrence versus the dimensionless time $\gamma_{0}T$ for vacuum Lorentzian environment with $\lambda=200\gamma_{0}$ and $\Delta=40\gamma_{0}$. The fidelity is calculated under the initial state $|\varphi(0)\rangle=(|0\rangle+|1\rangle)/\sqrt{2}$ of the atom.}
\end{figure}

In Fig.4, we show the evolutions of the fidelity and the normalized concurrence versus total dimensionless time for the vacuum Lorentzian environment, Where $\lambda=200\gamma_{0}$ and $\Delta=40\gamma_{0}$. In the calculation of fidelity, we again take the initial state of the atom to be $|\varphi(0)\rangle=(|0\rangle+|1\rangle)/\sqrt{2}$. It again shows the similar behaviors as that of Fig.2. In the limit of $\Delta t\rightarrow 0$, we have from Eq.(31) that $\mathbb{G}=[G(\Delta t)]^{T/\Delta t}\rightarrow 1$, so that the fidelity and the normalized concurrence approach to unity. The visible distinction of this figure is that there are some oscillations in the evolution (see black lines) when $\Delta t$ becomes larger, which is the result of non-Markovian effect.

\section{Markovian limit and quantum Zeno effect}

From the above discussion about dephasing and amplitude-damping SQC model, we find that if we break a noisy SQC into a series of sub-components, the quality of information transmission would be improved. However, two kinds of special cases need to be mentioned. One is the so-called Markovian limit. For the time-independent Markovian model of noise, the decay rate $\gamma_{p}(t)$ in Eq.(9) is a constant $\gamma_{p}^{M}$ so that $\Gamma(t)=\gamma_{p}^{M}t$. Therefore the evolution of Eq.(11) for the piecewise SQC is equivalent to that of an integral SQC with the same environment. Analogously, for the amplitude-damping SQC, the time-independent Markovian model of noise means that the decay rate $\gamma_{a}(t)$ and the Lamb shift $S(t)$ in Eq.(22) are replaced respectively by the constants $\gamma_{a}^{M}$ and $S^{M}$, so that $G(t)=\exp\{-(\gamma_{a}^{M}+iS^{M})t/2\}$. Thus Eq.(25) is also equivalent to the evolution of an integral SQC with the same environment. The idea for improving the quality of information transmission is obviously invalid in the Markovian limit.

We can even, in a more general sense, discuss the problem about Markovian limit. For the time-independent Markovian model of noise, the evolution of the atomic state is governed by the standard Lindblad master equation $\dot{\rho}=\mathcal{L}\rho$ with a time-independent generator $\mathcal{L}$. Such a master equation leads to a dynamical semigroup of one-parameter, completely positive and trace-preserving maps, $\varepsilon(t)=\exp(\mathcal{L}t)$. For the piecewise SQC in Fig.1, the whole process comprises $n$ independent sub-dynamics. Each sub-dynamics in itself evolves from zero time. If each sub-dynamics has identical generator $\mathcal{L}$, then the time translation invariance of the semigroup dynamics leads to $\rho(T)=\exp(\mathcal{L}T)\rho(0)$, i,e., the piecewise SQC is equivalent to an integral SQC with the same environment.
For the time-dependent Markovian or non-Markovian model of noises, the time translation invariance breaks and the result is invalid.

Another special case is the limit of $\Delta t\rightarrow 0$, which is related to the phenomenon of quantum Zeno effect \cite{Misra,Itano}. In the limit of $\Delta t\rightarrow 0$, the atom continually touches new environments, which is equivalent to the frequent measurements of the atomic state by the environments, making the atomic state unchanged. Also, this Zeno effect only could occur in the cases where the interaction of the atom with the SQC belongs to non-Markovian or time-dependent Markovian model. For the time-independent Markovian model of noise, the quantum Zeno effect is not to take place.

\section{Conclusion}
In conclusion, we have investigated the properties of a noisy SQC constituted by a series of identical segments, each segment modeled by a local environment. For the non-Markovian or time-dependent Markovian model of noise, we found when the length of each segment, i.e., $\Delta t$, keeps fixed, that the fidelity, the purity and concurrence of the output states decrease with the increasing of the SQC length (quantified by the time $T$), which agree with our intuitive senses. However, for the fixed length $T$ of the SQC, these physical quantities increase as $\Delta t$ gets shorter. It implies, by breaking a noisy SQC into a series of periodically arranged sub-components, that the quality of information transmission can be improved. In the limit of $\Delta t\rightarrow 0$, the quantum Zeno effect may be observed. For the time-independent Markovian model of noise, the discussed piecewise SQC is equivalent to an integral noisy SQC with the same environment. Thus the quantum Zeno phenomenon vanishes and the scheme for improving the quality of SQCs is invalid. Our results, on the one hand, present a possible method for improving the quality of quantum information transmission via noisy quantum channels. And on the other hand, they shed some light on the difference between time-dependent and time-independent Markovian processes.

In the discussion, the length of each segment was assumed to be identical. Nevertheless, the results are also valid for imperfectly periodic-structured SQCs. Of course, the results were derived only for the dephasing thermal bath or squeezed bath with Ohmic spectral density, or for dissipative Lorentzian environment. For other characteristic environments, further investigations are required.

A possible implementation of the scheme is worthy of mention. It is known that photons are the excellent carriers for quantum information transmission. Simultaneously, photonic crystal which exhibits explicit non-Markovianity \cite{Bellomo} is the good representative for realizing piecewise SQCs. By encoding quantum information into photons and let them transfer in photonic crystals, the quality of information transmission may be improved compared with the communication with optical fibers. We expect the demonstration of this scheme.

At the end, we point out that the idea for improving information transmission quality also applies to the information storage. In fact, if all the cavities in Fig.1 locate in the same place instead of forming a SQC, then the diagram corresponds to a process of information storage. Therefore by breaking a TQC into many shorter sub-processes, i.e., taking the information entity (qubit) constantly into new independent containers, the quality of information storage get improved if the interaction of the qubit with its environment belongs to the time-dependent Markovian or non-Markovian model of noises.

\section{ACKNOWLEDGMENTS}
This work is supported by the National Natural
Science Foundation of China (Grant Nos. 11275064,11075050), Specialized Research Fund for the Doctoral Program of Higher Education (Grant No. 20124306110003), the Program for Changjiang Scholars and Innovative Research Team in
University under Grant No. IRT0964, and the Construct Program of the
National Key Discipline.

\end{document}